\documentclass{article}

\usepackage{arxiv}

\usepackage{amsthm,amsmath}
\usepackage[utf8]{inputenc}
\usepackage[T1]{fontenc}
\usepackage{hyperref}
\usepackage{url}
\usepackage{booktabs}
\usepackage{amsfonts}
\usepackage{nicefrac}
\usepackage{microtype}
\usepackage{amssymb}
\usepackage{textgreek}
\usepackage{siunitx}
\usepackage{array}
\usepackage{tabu}
\usepackage{mathtools}
\usepackage{floatrow}
\usepackage{tabularx}
\usepackage{float}
\usepackage{algorithm2e}
\usepackage{listings}
\usepackage{bibentry}

\title{Neural Representations of Cryo-EM Maps and a Graph-Based Interpretation}

\author{
  Nathan Ranno\\
  Division of Computing and Software Systems\\
  University of Washington Bothell\\
  Bothell, WA 98011\\
  \texttt{nranno@uw.edu}\\
   \And
 Dong Si\thanks{Corresponding Author}\\
  Division of Computing and Software Systems\\
  University of Washington Bothell\\
  Bothell, WA 98011\\
  \texttt{dongsi@uw.edu}\\
}

\begin{document}
\maketitle

\begin{abstract}
Advances in imagery at atomic and near-atomic resolution, such as cryogenic electron microscopy (cryo-EM), have led to an influx of high resolution images of proteins and other macromolecular structures to data banks worldwide. Producing a protein structure from the discrete voxel grid data of cryo-EM maps involves interpolation into the continuous spatial domain. We present a novel data format called the neural cryo-EM map, which is formed from a set of neural networks that accurately parameterize cryo-EM maps and provide native, spatially continuous data for density and gradient. As a case study of this data format, we create graph-based interpretations of high resolution experimental cryo-EM maps. Normalized cryo-EM map values interpolated using the non-linear neural cryo-EM format are more accurate, consistently scoring less than 0.01 mean absolute error, than a conventional tri-linear interpolation, which scores up to 0.12 mean absolute error. Our graph-based interpretations of 115 experimental cryo-EM maps from 1.15 to 4.0 \si{\angstrom} resolution provide high coverage of the underlying amino acid residue locations, while accuracy of nodes is correlated with resolution. The nodes of graphs created from atomic resolution maps (higher than 1.6 \si{\angstrom}) provide greater than 99\% residue coverage as well as 85\% full atomic coverage with a mean of than 0.19 \si{\angstrom} root mean squared deviation (RMSD). Other graphs have a mean 84\% residue coverage with less specificity of the nodes due to experimental noise and differences of density context at lower resolutions. The fully continuous and differentiable nature of the neural cryo-EM map enables the adaptation of the voxel data to alternative data formats, such as a graph that characterizes the atomic locations of the underlying protein or macromolecular structure. Graphs created from atomic resolution maps are superior in finding atom locations and may serve as input to predictive residue classification and structure segmentation methods. This work may be generalized for transforming any 3D grid-based data format into non-linear, continuous, and differentiable format for the downstream geometric deep learning applications.
\end{abstract}


\section{Introduction}
Proteins serve an enormous amount of functions within organisms. Their functionality is prescribed by the form of their tertiary structure, which is the three-dimensional spatial arrangement of the composite amino acids. The sequence of amino acids that form the polypeptide chain, or the primary structure, ranges from tens to many hundreds of residues. Each primary structure is deterministic and, when folded into its native state, produces a unique tertiary structure. A viral capsid, for example, is composed of one or more repeating protein tertiary structures\cite{roosViralCapsidsMechanical2007}. The SARS-CoV-2 pandemic demonstrates the importance of modeling protein functions, as they relate to understanding the virus's interactions, propagation, drug treatment, and infection prevention via vaccines.

The field of 3D electron microscopy (3DEM) is fundamental to the determination and validation of protein structures. Traditional methods such as X-ray crystallography and nuclear magnetic resonance (NMR) spectroscopy have helped fill protein data banks with tens of thousands\cite{goodsellRCSBProteinData2020} of structures that are used in fields such as drug and vaccine development. Cryo-EM, a relatively newer single-particle technique for samples prepared at cryogenic temperatures\cite{thompsonIntroductionSamplePreparation2016}, has been shown as a source of high quality, high resolution structure maps\cite{nogalesDevelopmentCryoEMMainstream2016}. The recent improvements in data processing and computation speed have given cryo-EM the ability to capture atomic resolution\cite{nakaneSingleparticleCryoEMAtomic2020} and near-atomic resolution images of protein quarternary structures and other macromolecular structures. High resolution 3DEM is crucial to further solving and refining of protein structures.

Protein structure determination via computational methods are based on the readily available primary structure. They are faster than cryo-EM in producing a structure output, which may take many months per structure\cite{lyumkisChallengesOpportunitiesCryoEM2019}, however the complexity posed by large sequences and inter-woven structures is a limiting factor. Though methods are rapidly improving, even the state-of-the-art methods\cite{seniorImprovedProteinStructure2020, zhengDeeplearningContactmapGuided2019} as demonstrated in recent CASP competitions\cite{kryshtafovychCriticalAssessmentMethods2019} do not extend to predicting multi-domain structures. In contrast, cyro-EM imaging techniques observe structures in their natively folded state, providing the role of both structure determination and experimental validation.

Many tools exist to supplement the production of cryo-EM maps in the various stages of map development\cite{sulowayAutomatedMolecularMicroscopy2005, mastronardeAutomatedElectronMicroscope2005, zivanovNewToolsAutomated2018}. Once a map is produced, there still remains a non-trivial step of aligning the primary structure to density regions within the map\cite{scheresPreventionOverfittingCryoEM2012}. Despite the imaging improvements, experimental maps often contain noise and other artifacts that lead to a time-consuming and manual structure determination process supported by cryo-EM maps visualization tools\cite{pettersenUCSFChimeraVisualization2004, emsleyFeaturesDevelopmentCoot2010}. A number of solutions exist for the partial\cite{frenzRosettaESSamplingStrategy2017, terashiNovoMainchainModeling2018} and full\cite{terwilligerFullyAutomaticMethod2018} automation of this process. Deep learning has also demonstrably improved both the automation and execution time of producing predicted protein structures in cryo-EM maps\cite{siDeepLearningPredict2020, pfabDeepTracerFastNovo2021}.

Fundamental to the operation of these tools is the transition from the 3D grid-aligned voxel data format of the cryo-EM map to a continuous spatial coordinate system. Simply labeling the original voxels with atomic types will not produce an accurate tertiary structure prediction as the native size of high resolution cryo-EM voxels generally range from 0.5 \si{\angstrom} to 1.5 \si{\angstrom}. With high resolution cryo-EM maps, we assume that areas of high density are indicative of atomic locations. Existing prediction methods use various amounts of sub-sampling the maps with linear interpolation and averaging of density values to determine atomic locations. However these types of interpolation cannot globally produce both non-grid points as well as density maximums, for the calculated maximums lie on the voxel grid. Non-linear interpolation methods are necessary in order to calculate density maximums that lie off the voxel grid.

Graphs are an intriguing data format for proteins and other molecular structures due to their physical similarity to the underlying data, and compared to cryo-EM images, the graph data format is much more condensed and concise. Graph-based methods and graph convolutional networks (GCN)\cite{kipfSemiSupervisedClassificationGraph2017, danelSpatialGraphConvolutional2020} are gaining popularity for tasks related to proteins, such as protein-protein interaction (PPI)\cite{yangGraphbasedPredictionProteinprotein2020, xiaoGraphEmbeddingbasedNovel2020}, protein function classification\cite{zamora-resendizStructuralLearningProteins2019, gligorijevicStructureBasedProteinFunction2020}, and primary structure alignment onto tertiary structures\cite{strokachFastFlexibleProtein2020, liSequenceguidedProteinStructure2020}. Creating a graph from the cryo-EM format relies on the ability to find dense points in the map, however, the variation of density values for similar atoms and the presence of noise in experimental maps are challenging for voxel thresholding techniques (Figure \ref{fig:threshold_problems}a).

In this paper, we present a novel data format for high-resolution cryo-EM maps that can produce a fully continuous, non-linear interpolation of the EM data using neural network representation. Namely, the SIREN architecture\cite{sitzmannImplicitNeuralRepresentations2020a} provides the basis for the interpolation. Our implementation automatically converts native 3D array data to the so-called neural cyro-EM maps and retains the ability to accurately reproduce the original input. This format may be extended in many ways, and as a case study, we create a novel graph-based interpretation of cryo-EM maps based on the neural network representation on the basis that there is a correlation between atomic locations to points of high density within the cryo-EM map. We show that the graph coverage of the cryo-EM data and node placement is well-suited for additional predictive methods to determine molecular structure.

\section{Methods}
The basis for our continuous and non-linear interpolation of cryo-EM maps is the use of one or more neural networks that parameterize the underlying map, producing EM density values given spatial coordinates. Though there is an extremely diverse set of possible neural network architectures for this task, the SIREN network architecture has been shown superior to other network types in ability to accurately represent natural signals\cite{sitzmannImplicitNeuralRepresentations2020a}. The sine layer presented by Sitzmann et al. uses, along with a specific weight initialization scheme, a sine function to wrap a linear transformation of the vector $\textbf{x}$, with a weight matrix $\textbf{W}$ and biases $b$: $y=\sin(\textbf{W}\textbf{x} + b)$. The success of SIRENs in preserving the derivatives of the original signals is also of particular note, underlying the architecture's performance in interpolation tasks. We propose that extending the architecture to three-dimensional images would show success in signal representation and interpolation.

\subsection{Pre-processing}
Cryo-EM maps are deposited and stored in the EM databases as three-dimensional arrays, where each index contains a density value. The array axes, commonly referred with the labels \textit{i}, \textit{j}, and \textit{k}, are correlated to real spatial coordinates by their respective voxel size in that axis. Each map contains a header that describes voxel sizes, which may not be uniform in each axis, as well as the relationship of the \textit{i}, \textit{j}, and \textit{k} axes to the spatial coordinate axes, which we label \textit{x}, \textit{y}, and \textit{z}. Cryo-EM maps may not be consistent with other maps in terms of the arrangements of axes in relation to spatial coordinates. We account for this by maintaining that the SIREN network inputs are the \textit{x}, \textit{y}, and \textit{z} coordinates respectively and transposing the voxel data to be consistent with this arrangement.

The density values of cryo-EM maps are unit-less and are not consistent between maps. Therefore we apply a normalization to the values across the entire map to the range of [-1, 1] which is suitable for training SIRENs. This is a two-step process. The first pass of normalization sets the negative density values of the original map to the lower bound of zero and scales the remaining values to the range of [0, 1].

\[
d_1 = 
\begin{cases} 
  0 & d_0 \leq 0 \\
  \frac{d_0 - d_{min}}{d_{max} - d_{min}} & d_0 > 0
\end{cases}
\]

The second pass expands the range to [-1, 1].

\[
d = 2d_1 - 1
\]

Initially, the variation in cryo-EM map shape led to the approach of varying the neural network size accordingly. However, after experimenting with the parameters for scaling the networks, the range of cryo-EM map sizes proved to be too large to effectively scale the networks. Even medium-sized maps took an incredibly long time in the SIREN training stage, due to both the size of the network and number of voxels used in training. By using a static neural network architecture across one or more sub-regions of the cryo-EM map, the training time of the entire map scales linearly with the voxel count. This approach also leads to a straightforward multi-GPU strategy when creating neural representations for cryo-EM maps that require multiple SIREN networks.

The voxel data of cryo-EM maps are divided into three-dimensional sub-regions of voxels with the maximum length of any axis of the sub-region being limited to 64 voxels. Each sub-region overlaps with its neighboring sub-regions along each coordinate axis by no less than four voxels. If the total number of voxels in a given axis is $n_v$, the starting index for each region $i_n$, is calculated using the region size $v_r$, the number of regions $n_r$, and the spacing interval $s$.
\[
v_r =
\begin{cases}
  n_v & n_v < 64 \\
  64 & n_v \geq 64
\end{cases}
\]
\[
n_r = \lceil \frac{n_v - v_r}{v_r - 4} \rceil + 1
\]
\[
s = \frac{n_v - v_r}{\max(n_r - 1, 1)}
\]
\[
i_n = \lfloor (n - 1)s \rfloor, n \in \{\mathbb{Z} \mid 1 \leq n \leq n_r\}
\]
These calculations are performed per axis, and the results are used to slice the voxel data into sub-regions. The sub-region voxels are used to train distinct SIRENs, and the boundaries are used in querying density and gradient values, as shown in Equations \ref{eq:nominald} and \ref{eq:offnominald}.

\subsection{Training}
\begin{figure}
    \includegraphics[width=0.6\linewidth]{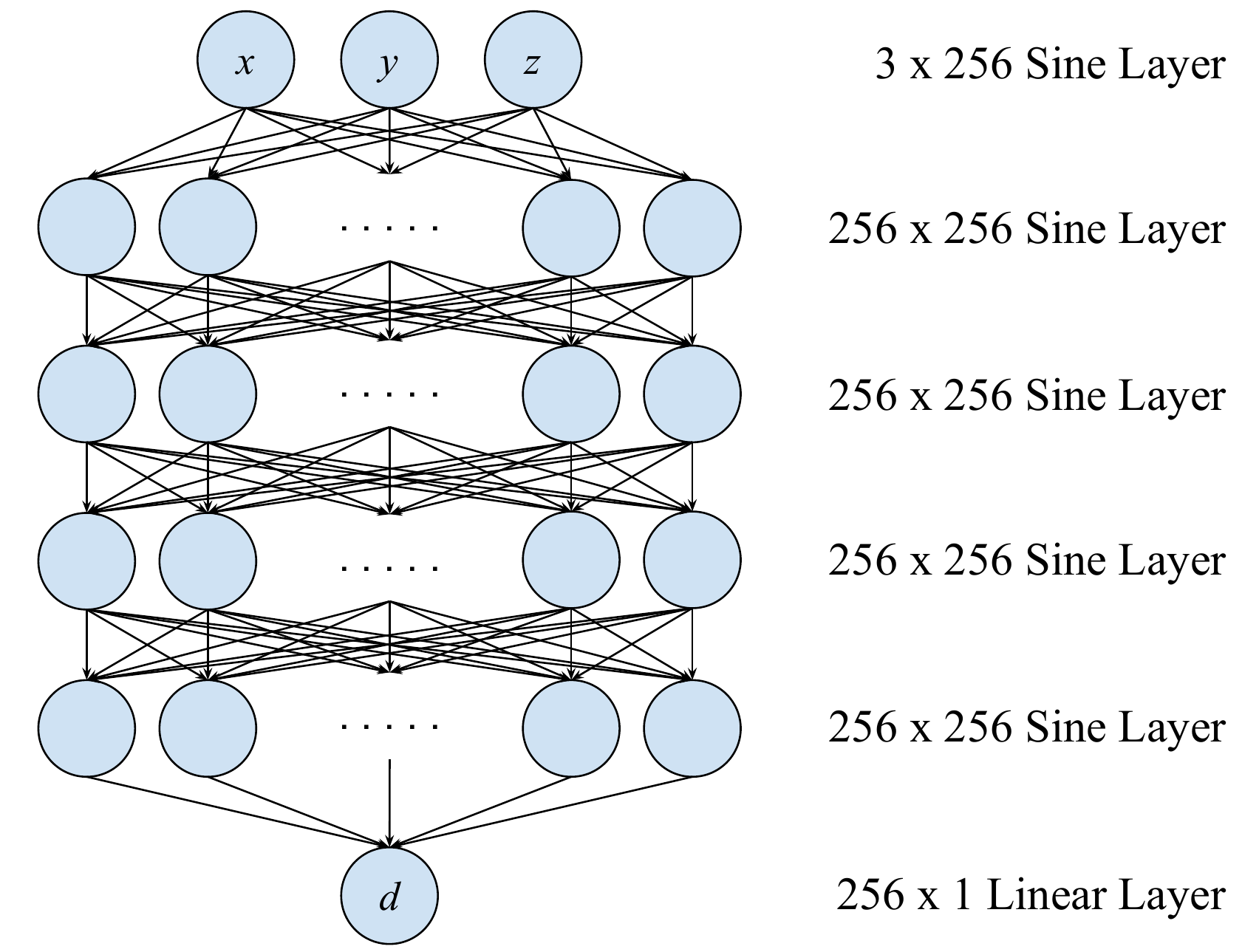}
    \caption{Architecture of each SIREN used to represent a region of voxels. There are 256 hidden features per 4 hidden layers with a final linear layer to output the cryo-EM density given an spatial coordinate.}
    \label{fig:siren_architecture}
\end{figure}

Cryo-EM maps are pre-processed and divided into overlapping sub-regions with a maximum size of 64 voxels in each axis, where each sub-region is allocated a distinct SIREN neural network. The network architecture (Figure \ref{fig:siren_architecture}) is a fully connected multi-layer perceptron with a sinusoidal input layer, four hidden sinusoidal layers with 256 features, and a final linear output layer. Weights are initialized in the network from uniform distributions $\mathcal{U}(-\sqrt{6/n}, \sqrt{6/n})$ for the first layer and $\mathcal{U}(-(1/\omega_0)\sqrt{6/n}, (1/\omega_0)\sqrt{6/n})$ for the subsequent layers, with $n=256$ and $\omega_0=30$. With three input dimensions and one output dimension, the network fully loaded with the maximum sub-region size results in roughly 4 GB of memory space, which may be accommodated by most GPUs commonly used for deep learning. The network architecture and sub-region size were determined to balance spatial coverage, training time, and non-volatile storage space. We use the PyTorch\cite{NEURIPS2019_9015} framework for neural network and many data manipulation operations.

The SIREN training process utilizes the observed periodic behavior of the network fit improving for many epochs and then briefly regressing, leading to an overall network fit improvement for each of these cycles. Using the mean squared error (MSE) as the loss function along with the Adam optimizer\cite{kingmaAdamMethodStochastic2017}, the network is trained to its ``natural fit point'', which we define as the point at which the lowest MSE loss value, $l_{min}$, has not been improved for 25 epochs while $0.00001 \leq l_{min} < 0.0004$. If $l_{min} < 0.00001$, the training loop exits immediately. In practice, this reduces the noise ceiling for cryo-EM regions with relatively smoother contents, such as empty space. Since the goal of SIRENs is to fit the network to all the voxels in the map, there are no separate data splits for validation and testing. The separate nature of each SIREN leads to the ability to parallelize the training of a neural cryo-EM map over multiple GPUs.

\subsection{Data Retrieval}
When allocating spatial coordinates to sub-regions, we observed discontinuities in the output values when using hard boundaries between sub-regions. Therefore, we employ a strategy of weighing the output of the distinct networks by the coordinate's position along the axes of overlap and performing an average with the output of all networks that contain the coordinate. Once the SIRENs are trained, the sub-regions are patched together such that any spatial coordinate $p$ in the cryo-EM map exists at a point where sub-regions overlap along $n$ axes, and given that sub-regions were created to overlap strictly along the coordinate axes, the possible values of $n$ are 0, 1, 2, and 3. The nominal case, $n = 0$, means only one neural network $r_0$ is used to produce the output $d$:
\begin{equation}\label{eq:nominald}
    d = r_0(p)
\end{equation}
In the off-nominal cases, $1 \leq n \leq 3$, for each axis overlap there are two regions $r_{n1}$ and $r_{n2}$ that produce output at the coordinate. The network outputs are weighted by the corresponding coordinate component's distance along the overlap and averaged with any other overlapping axes:
\[ w_{n1} = \frac{r_{n1end} - p_n}{r_{n1end} - r_{n2start}}, w_{n2} = 1 - w_{n1} \]
\begin{equation}\label{eq:offnominald}
    d = \frac{\sum_{n=1}^{3} w_{n1}r_{n1}(p) + w_{n2}r_{n2}(p)}{n}
\end{equation}

The density value is query-able at any point in the neural cryo-EM map. To allow for meaningful human inspection and rendering of SIREN output in common 3DEM tools such as Chimera\cite{pettersenUCSFChimeraVisualization2004} and Coot\cite{emsleyFeaturesDevelopmentCoot2010}, we transform a query's SIREN output, $d_out$, from the internal nominal range of [-1, 1] to the nominal range of [0, 1] to give the final output $d$.
\[
d = \frac{d_{out} + 1}{2}
\]
The actual values, $d_{out}$, are not strictly constrained to the range of [0, 1] in order to allow the network to interpolate beyond the range of the initial training data.

While the neural cryo-EM data format provides the ability to sample density data from anywhere in the map, it is important to distinguish that it does not increase the data resolution. The data format provides a non-linear interpolation of density that is fully continuous and differentiable. This format may be employed and extended in different ways, but for this paper we present a novel graph-based format constructed using the neural cryo-EM map.

\begin{figure}
    \includegraphics[width=\linewidth]{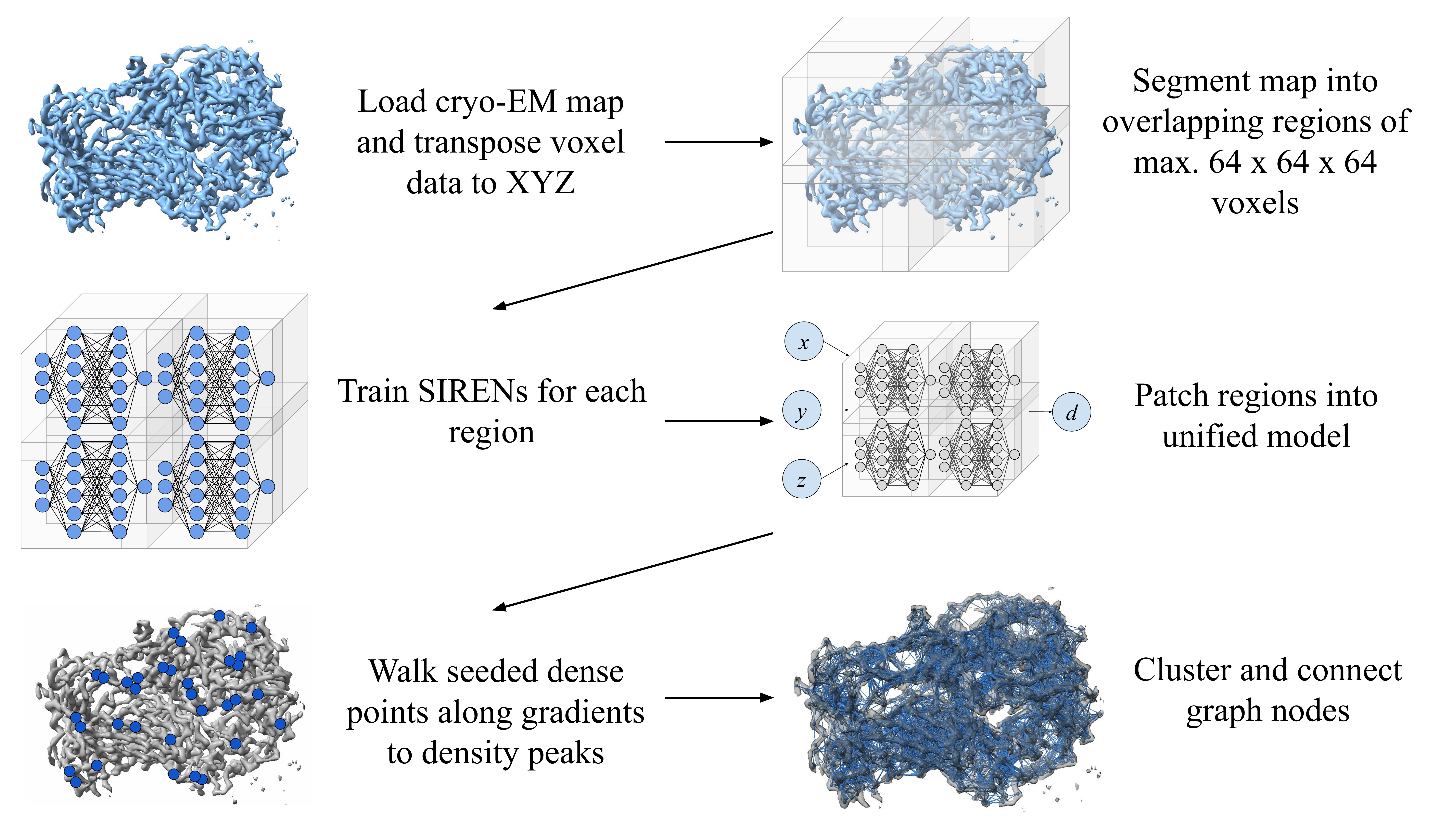}
    \caption{Overview of the creation process of the density map graph. 1) Transpose, if necessary, voxel data in the map to align with a consistent XYZ view. 2) Divide the map into regions of max. 64x64x64 voxels that overlap by no less than four voxels in each axis. 3) Train a SIREN for each region using one or more GPUs in parallel. 4) Patch the regions together such that any input coordinate produces an output by Equations \ref{eq:nominald} and \ref{eq:offnominald}. 5) Use the neural cryo-EM format to seed the spatial area with points and iteratively walk the points along their gradients to density peaks. 6) Cluster the points into nodes and connect them based on an adjacency threshold.}
    \label{fig:full_density_graph_creation}
\end{figure}

\subsection{Cryo-EM Density Graphs}
Our graph-based interpretation of cryo-EM maps is an extension of the neural cryo-EM map format. The nodes of the graph describe the locally dense spatial coordinates throughout the map. The context of the nodes in relation to the molecular structure depends on the resolution of the underlying cryo-EM map. In general, for high resolution maps, the intention is for nodes to correspond to amino acid residue locations. The lack of cryo-EM map annotation means that the resulting feature vector of each node is composed of four values: the three-dimensional spatial coordinate and the density at that location. Each edge of the graph may contain an optional feature vector, depending on the intended use of the graph in a future downstream implementation. The edge's feature vector is a single dimension with the value of the edge's length. See Figure \ref{fig:full_density_graph_creation} for an visualization of the complete density graph creation process.

The method by which the graphs are created relies heavily on the fully continuous and differentiable nature of the neural cryo-EM representation. At any spatial coordinate contained in the cryo-EM map, the neural format may be queried for a density value and a gradient vector, which gives the magnitude and direction of density increase. The graph creation is the ensemble of summarily naive steps which are made possible by the neural cryo-EM format. We employ a global generic thresholding mechanism to filter out irrelevant regions of the cryo-EM map. When creating the neural cryo-EM map, the mean and standard deviation of the normalized SIREN training data are retained. The threshold value $T$ is given as $T=\mu + 3\sigma$, and seed points are determined by sampling the entire map with a $0.5$ \si{\angstrom} step size in each axis, discarding points below the value of $T$.

Each seed point is iteratively moved along its gradient vector until a density peak is detected (Algorithm \ref{alg:gradient-walk}). The density peak is the point of highest detected density while traversing the gradient vector with a step size of $0.05$ \si{\angstrom}. For each step iteration, the point location and density are retained for reference against the next step in order to compare densities and provide the correct position in the results. It is possible that traversing in the direction of the gradient results in points exiting the spatial domain of the map. In this case, the seed point is simply removed from the pool of seed points.

\begin{algorithm}[h!]
    \KwData{Seed points of the map}
    \KwResult{All seed points either deleted or moved to a density peak.}
    \While{seed points remain in pool}
    {
        \For{each point in the pool}
        {
            \If{point not in map}
            {
                remove point from pool\;
            }
            calculate density and gradient vector\;
            \eIf{calculated density $\leq$ cached density}
            {
                add cached point location to results\;
                remove point from pool\;
            }
            {
                cache current position\;
                cache current density for point\;
                modify point location by 0.05 \si{\angstrom} in direction of gradient vector\;
            }
        }
     }
     return results\;
     \caption{Gradient-walking algorithm used in detecting local density peaks using the neural cryo-EM map format}
     \label{alg:gradient-walk}
\end{algorithm}

Once all density peaks have been reached, the DBSCAN\cite{esterDensitybasedAlgorithmDiscovering1996} clustering algorithm is performed on the points. For our case, two points are considered part of the same cluster if they are within $0.2$ \si{\angstrom} of the other point, and points without a neighbor within that range are considered their own cluster. While this latter setting decreases the specificity of the node placement and makes it more susceptible to noise, we found that the tradeoff with an increase of overall sensitivity was worth it. The centroid is calculated for each cluster as a potential graph node.

While our graph evaluation is focused on the node placement and representation of the underlying deposited structures, graphs are not simply a set of nodes. We connect nodes with edges based on an arbitrary spatial adjacency threshold 2 \si{\angstrom} times greater than the reported resolution of the underlying cryo-EM map. In order to create the edges, a pairwise adjacency matrix is computed over all nodes, and any indices whose value is below the threshold are used to create node pairs. The remaining nodes and edges compose the output graph-based interpretation of a cryo-EM map.

\subsection{Data Sets}
This paper derives two sets of cryo-EM data, one of simulated maps and the other of experimental maps, from the same source. The base for the datasets is all the high resolution ($\leq 4$ \si{\angstrom}) cryo-EM maps from the EM Data Resource\cite{lawsonEMDataBankOrgUnified2011} that have an associated PDB deposition. Each deposition is often one of many from a given publication, and we filter the base dataset for only the map and corresponding structure of highest resolution from each publication. This significantly reduces the number of duplicate and very similar entries in the data pool.

\begin{figure}
    \includegraphics[width=0.6\linewidth]{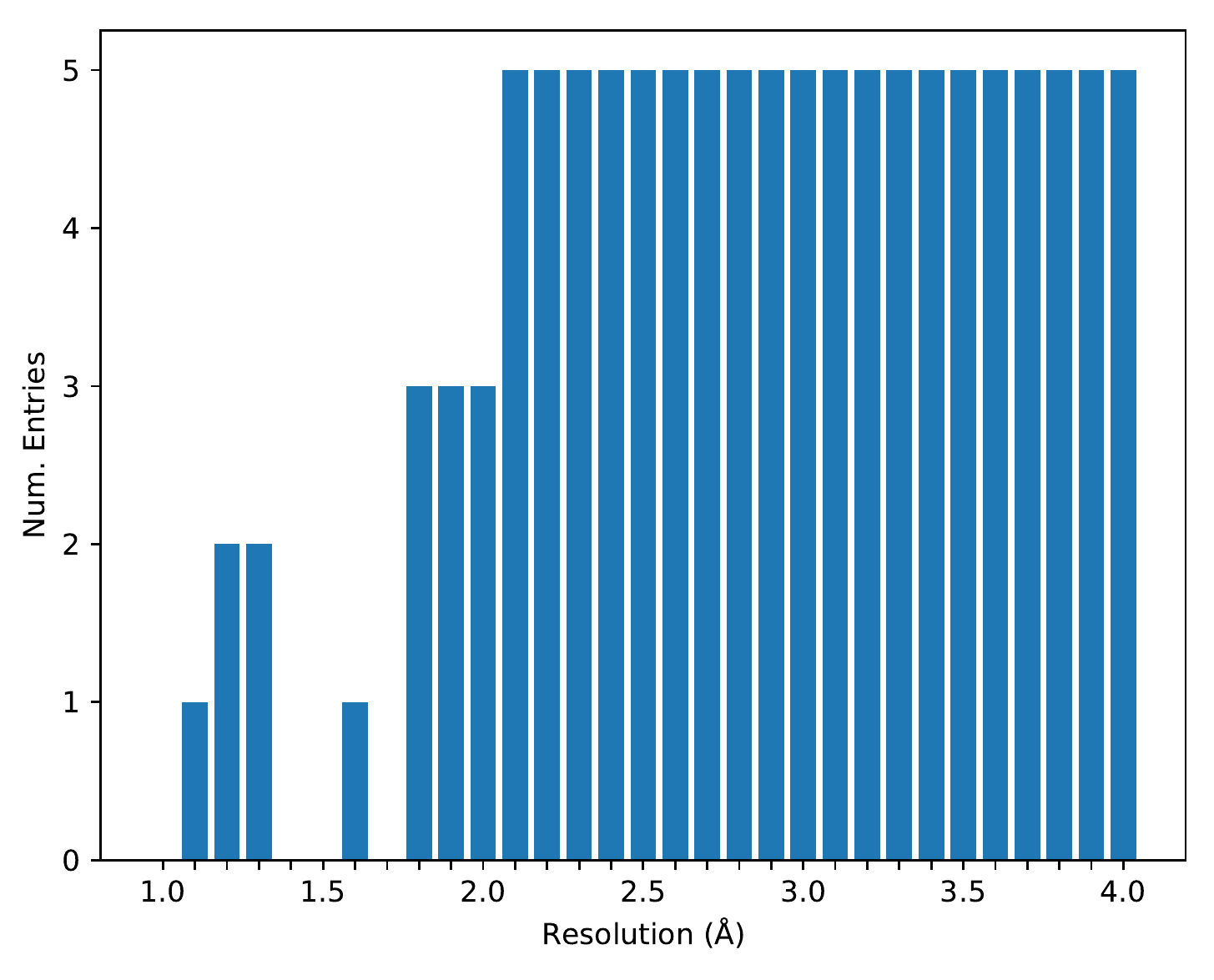}
    \caption{The distribution of the data set across resolution. There are unrepresented resolutions and a general lack of maps at the high end of the resolution range.}
    \label{fig:dataset_dist}
\end{figure}

From this data pool, entries are grouped by resolution rounded to the nearest $0.1$ \si{\angstrom}. Up to five entries are randomly selected from each group to represent that slice of the resolution range. If five maps entries are not available in a group, then all the entries of the group are selected. Additionally, if a cryo-EM map of a selected entry contains more than $512^3$ voxels, the candidate entry is ignored, and the random selection is retried with that map removed from the pool. This results in a total of 115 high resolution cyro-EM maps and corresponding deposited structures. Figure \ref{fig:dataset_dist} shows the distribution of the data across the range of resolutions.

\section{Results}
A dataset of 115 cryo-EM maps, both experimental and simulated, and corresponding PDB-deposited\cite{goodsellRCSBProteinData2020} structures across the range of high resolutions ($\leq 4$ \si{\angstrom}) serves as the basis for the experiments presented in this paper. By constructing simulated maps from the deposited structures at the reported experimental cryo-EM resolutions, the neural cryo-EM map's capability for interpolation is evaluated, comparing it against a tri-linear interpolation. Experimental cryo-EM maps are used to present a novel use for this data format, a graph-based interpretation of cryo-EM maps. We evaluate the graphs' coverage of the underlying structure along with the accuracy of node placement with respect to residue and atom locations. The results are also compared to the prediction output of the state-of-the-art cryo-EM modelling tool DeepTracer\cite{pfabDeepTracerFastNovo2021}.

\begin{figure}
    \includegraphics[width=\linewidth]{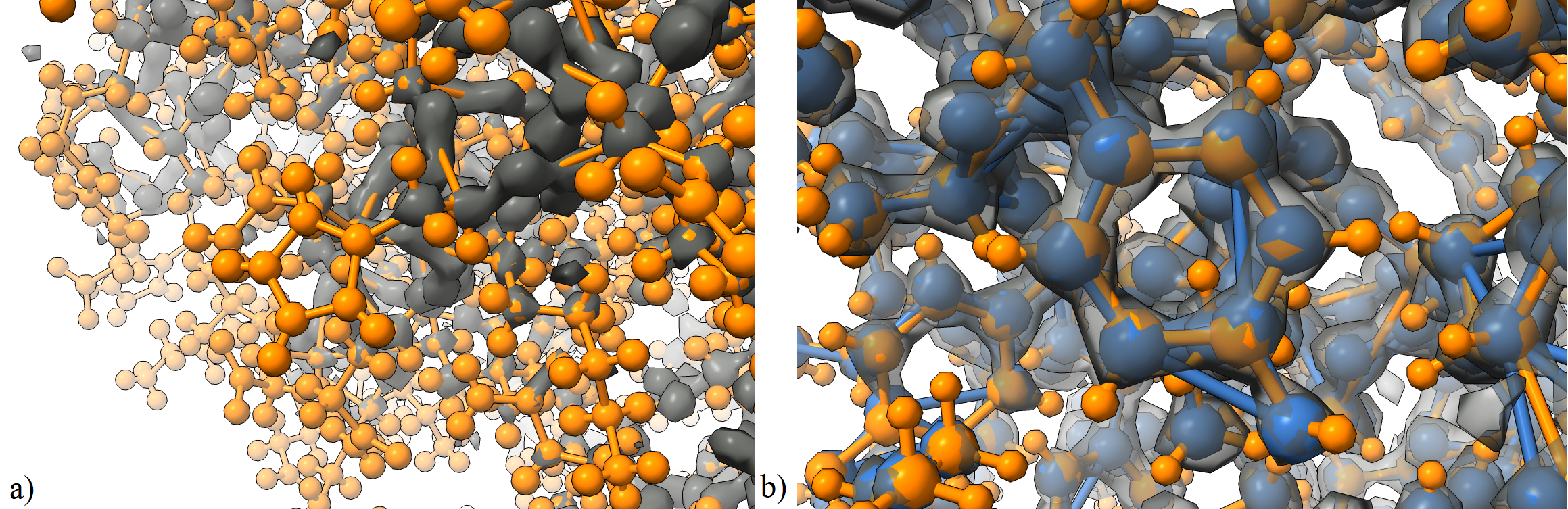}
    \caption{a) A threshold high enough to distinguish separate clusters of voxels is too high for many side chain atoms and atoms on the periphery of the protein structure (orange), demonstrated using EMD-11103. The dark gray voxel data is not uniformly representative of all atoms given a high-pass filter value, which illustrates the need for a better representation of the map. b) Localized view of graph nodes (blue) created by our method for the atomic resolution map EMD-11103 human apoferritin (transparent gray) and the atom locations of the corresponding deposited structure PDB-6z6u (orange).}
    \label{fig:threshold_problems}
\end{figure}

\subsection{Non-linear Interpolation}
Simulated cryo-EM maps, created using Chimera's \textit{molmap} tool\cite{pettersenUCSFChimeraVisualization2004}, are constructed from the sum of resolution-dependent Gaussian functions centered on atomic locations, providing determinism throughout the spatial region of the map. This determinism acts as the control in the evaluation of the interpolation capability of the neural cryo-EM map format. For each deposited structure in the 115-map dataset, the reported resolution of the corresponding cryo-EM map is used as the target resolution for \textit{molmap}. Two simulated maps are created per entry, one with the default voxel size of $(resolution / 3)$ to serve as the input to experimental interpolators and the other with a voxel size of $0.2$ \si{\angstrom} to serve as the control for interpolated values. All other arguments remain their default value.

We evaluate the interpolations against the control by the metric of mean absolute error (MAE) with the initial and control maps' voxel values normalized to the range of [0, 1]. The neural cryo-EM map format, by its nature, contains a very low level of global background noise correlated to the amount of loss existing at the exit point of the training loop. This noise is absent in a tri-linear interpolation, again by its nature. To reduce the influence of the known neural noise, only the voxels of the control that contain a non-zero value are used in the comparisons of the types of interpolations.

\begin{figure}
    \includegraphics[width=0.6\linewidth]{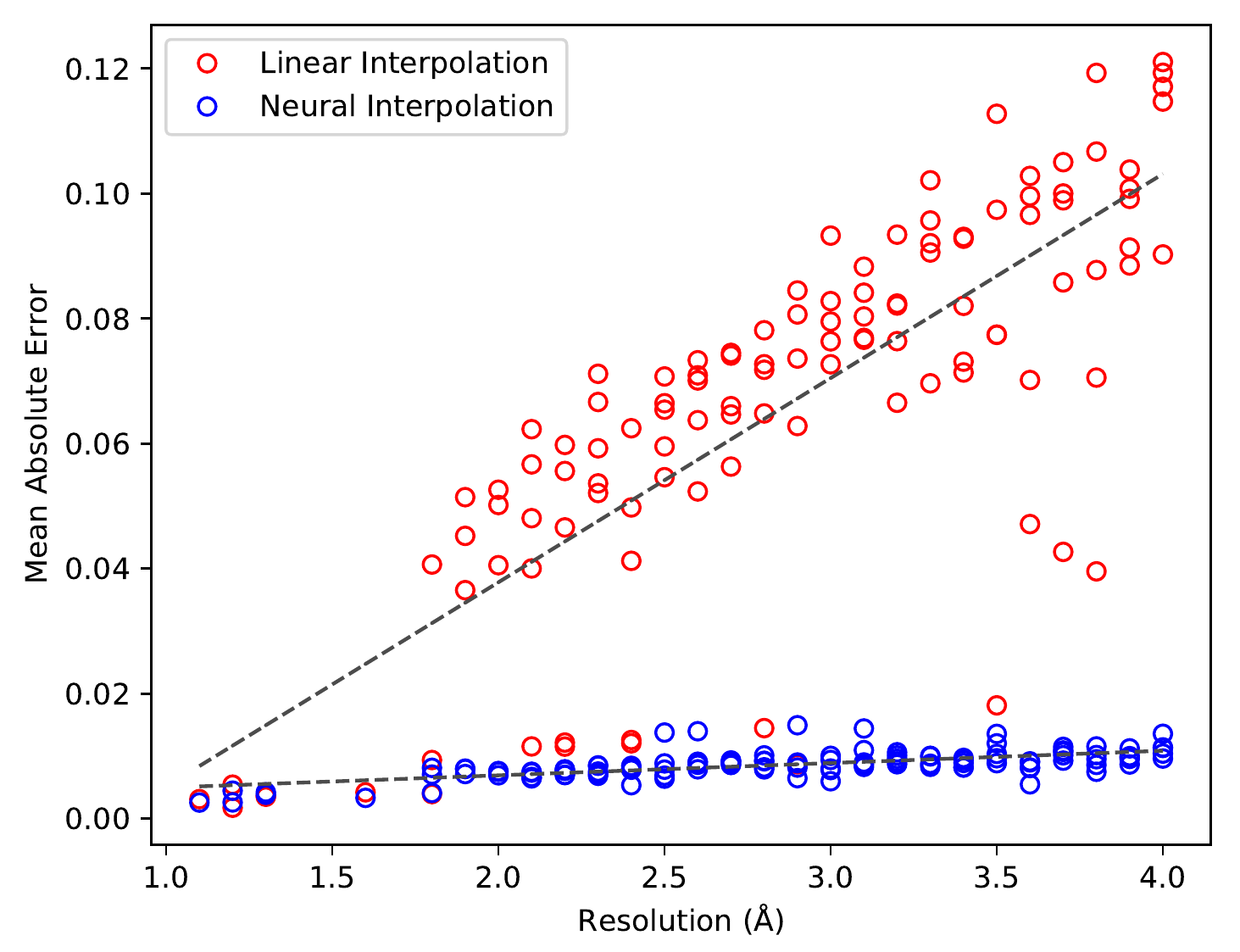}
    \caption{Plot comparing the mean absolute error of linear and the neural interpolation against 115 simulated cryo-EM maps. The neural interpolation significantly outperforms the linear interpolation and does not suffer worse performance as resolution increases.}
    \label{fig:interpolation_compare}
\end{figure}

As shown in Figure \ref{fig:interpolation_compare}, the interpolation performance of the neural cryo-EM map format is an order of magnitude more accurate to the true values of simulated map, consistently producing a MAE of $< 0.01$, and it does not appear to worsen with a decrease in resolution. The tri-linear interpolation not only worsens with decreasing resolution, it is unable to capture non-linearity of the underlying data, effectively performing a smoothing of density peaks in the map. The overall average MAE for the linear interpolation is $0.066$, but rises to $0.12$ for the lowest resolutions in the range. In contrast, the learned neural representations of the maps are able to not only capture the non-linearity but also preserve the ability to interpolate density values higher or lower than the initial maps' inputs.

In very high resolution simulated maps, the interpolation performance is exceptionally high at about $0.0005$ MAE, but the gap in performance between the two interpolators is not pronounced. This is likely due to the rapidly diminishing Gaussian function around atomic locations of the underlying control map, which means that the original simulated map has already captured much of the density data. The amount of interpolation performed is also less in these cases due to the initially small voxel sizes on those maps. For example, an atomic-level resolution map simulated at $1.2$ \si{\angstrom} has a voxel size of $0.4$ \si{\angstrom}, meaning only $2$ voxels are interpolated for every voxel of the original source.

As shown, the neural cryo-EM map format contains the ability to capture the non-linearity of underlying data, preserve density peaks, and provide spatially continuous and differentiable data. This leads to many extensions of the data format beyond what the conventional voxel representation provides. One such format is a graph, which we present in the next section.

\subsection{Graph-based Interpretation}
A density graph is created from a neural cryo-EM map by seeding the map with points, incrementally adjusting the points' position in the direction of their gradient vectors and stopping when a peak is detected, clustering the points using the DBSCAN\cite{esterDensitybasedAlgorithmDiscovering1996} algorithm, and calculating the centroid of each cluster. Candidate nodes are placed at the centroid locations and connected with edges based on an adjacency threshold and sub-graph constraints using the NetworkX\cite{SciPyProceedings_11} library.

From our dataset of 115 experimental cryo-EM maps, there are six maps at or below 1.6 \si{\angstrom} resolution, which we consider ``atomic resolution.'' The dense points in these maps are largely correlated to individual atoms, as opposed to the general location of the amino acid residues. Table \ref{tab:atomic_eval} shows the evaluation the cryo-EM density graphs from atomic resolution maps against all atoms documented in their respective deposited structures. The constraint in the sensitivity and specificity calculations is a 1 \si{\angstrom} radius around atoms and nodes. Considering just the C-$\alpha$ atoms, typically the most prominent atoms in amino acid residues, the density graphs provide a coverage of $99.4\%$. For all atoms documented in the PDB-deposited structure, the density graphs provide an average $85.2\%$ coverage as well as an average $88.4\%$ specificity rating of nodes to atoms. The RMSD of the matching nodes to their respective atomic locations is very low with a mean value of 0.19 \si{\angstrom} across the atomic resolution density graph set. Figure \ref{fig:threshold_problems}b depicts an atomic resolution density graph compared with the deposited structure and the voxel grid data.

\renewcommand{\tabularxcolumn}[1]{m{#1}}
\begin{table}
\begin{tabularx}{0.9\textwidth}{|c|c|c|c|>{\centering\arraybackslash}X|>{\centering\arraybackslash}X|} 
    \hline
    \multicolumn{6}{|c|}{Atomic Resolution Density Graphs} \\
    \hline
    EMDB ID & PDB ID & Resolution (\si{\angstrom}) & RMSD (\si{\angstrom}) & Sensitivity\newline (1 \si{\angstrom}) & Specificity\newline (1 \si{\angstrom}) \\ 
    \hline
    11668 & 7a6a & 1.15 & 0.138 &  87.98 &  96.21 \\
    11638 & 7a4m & 1.22 & 0.192 &  87.00 &  88.81 \\
    11103 & 6z6u & 1.25 & 0.147 &  88.73 &  95.34 \\
    11669 & 7a6b & 1.33 & 0.151 &  84.25 &  94.32 \\
    22657 & 7k3v & 1.34 & 0.226 &  78.82 &  89.44 \\
    11121 & 6z9e & 1.55 & 0.291 &  84.44 &  66.28 \\
    \hline
\end{tabularx}
\caption{Evaluation of cryo-EM density graph node locations against all atomic locations contained in the PDB-deposited structure file. The sensitivity is the percentage of atoms that match exclusively with a node within 1 \si{\angstrom}. The specificity is the percentage of nodes that match exclusively with an atom within 1 \si{\angstrom}.}
\label{tab:atomic_eval}
\end{table}

Due to the nature of the method as essentially a ``dense point detector,'' the context of what a node might represent differs between cryo-EM resolutions. In the $< 1.6 \si{\angstrom}$ range, dense points largely represent individual atoms of the protein residues, including the atoms in the both the backbone and side chains. With other high resolution maps $\leq 4 \si{\angstrom}$, the dense points are more indicative of amino acid residue locations, but are not precisely atomic locations. Despite this, the C-$\alpha$ atom locations of the deposited structure serve as the basis for evaluating the graph's of near-atomic resolution maps because every amino acid residue contains one, and they are relatively centrally located in a given residue. At the near-atomic resolution, nodes and residues are considered matched if a node is within 3 \si{\angstrom} of the C-$\alpha$ atom. 

\begin{figure}
    \includegraphics[width=\linewidth]{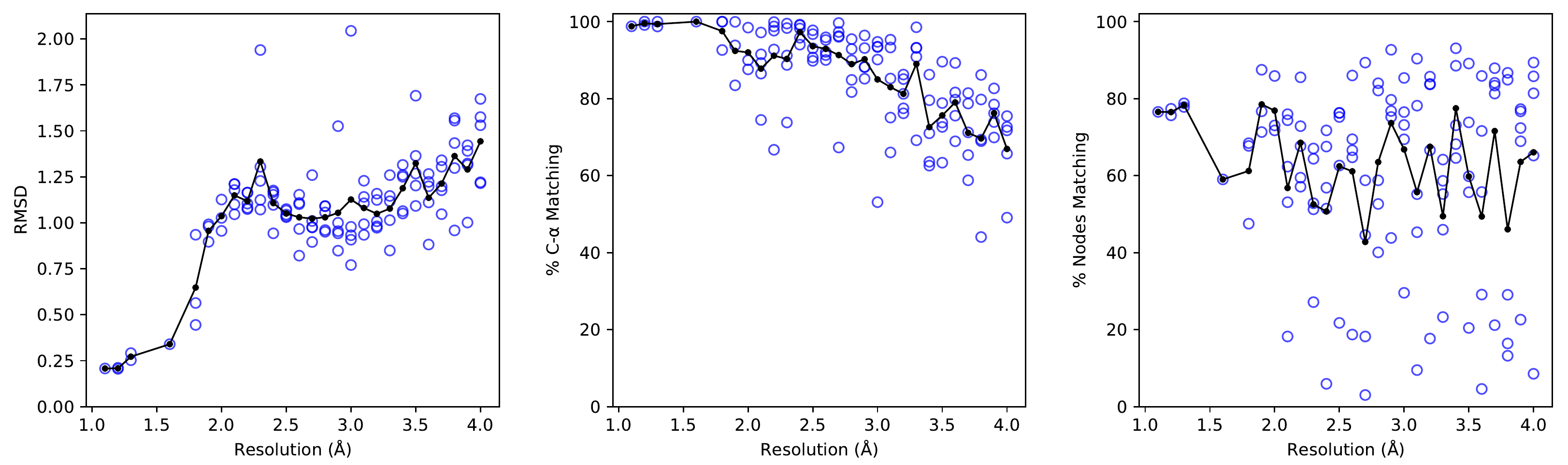}
    \caption{For all experimental maps in the 115-map dataset: the root mean square deviation (RMSD) of the closest node to C-$\alpha$ atoms (left), the percentage of C-$\alpha$ atoms of the deposited structure that have a node within 3 \si{\angstrom} (middle), and the percentage of total nodes that are within 3 \si{\angstrom} of a C-$\alpha$ atom in the deposited structure (right). Superimposed on the scatter graphs is the average value across all points at the given resolution in 0.1 \si{\angstrom} increments.}
    \label{fig:density_graph_results}
\end{figure}

Our results (Figure \ref{fig:density_graph_results}) show a clear delineation along the boundary of atomic and near-atomic resolutions. The near-atomic resolution graphs ($> 1.6 \si{\angstrom}$) have a mean RMSD of $1.13$ \si{\angstrom} and $84.5\%$ match. As resolution increases, the graphs' performance in these metrics decreases, which is expected. Interestingly, a jump in RMSD values appears between the atomic and near-atomic resolution maps, suggesting the possibility that in the latter maps the most dense point in a local area does not correspond exactly to the C-$\alpha$ location.

\begin{figure}
    \includegraphics[width=0.8\linewidth]{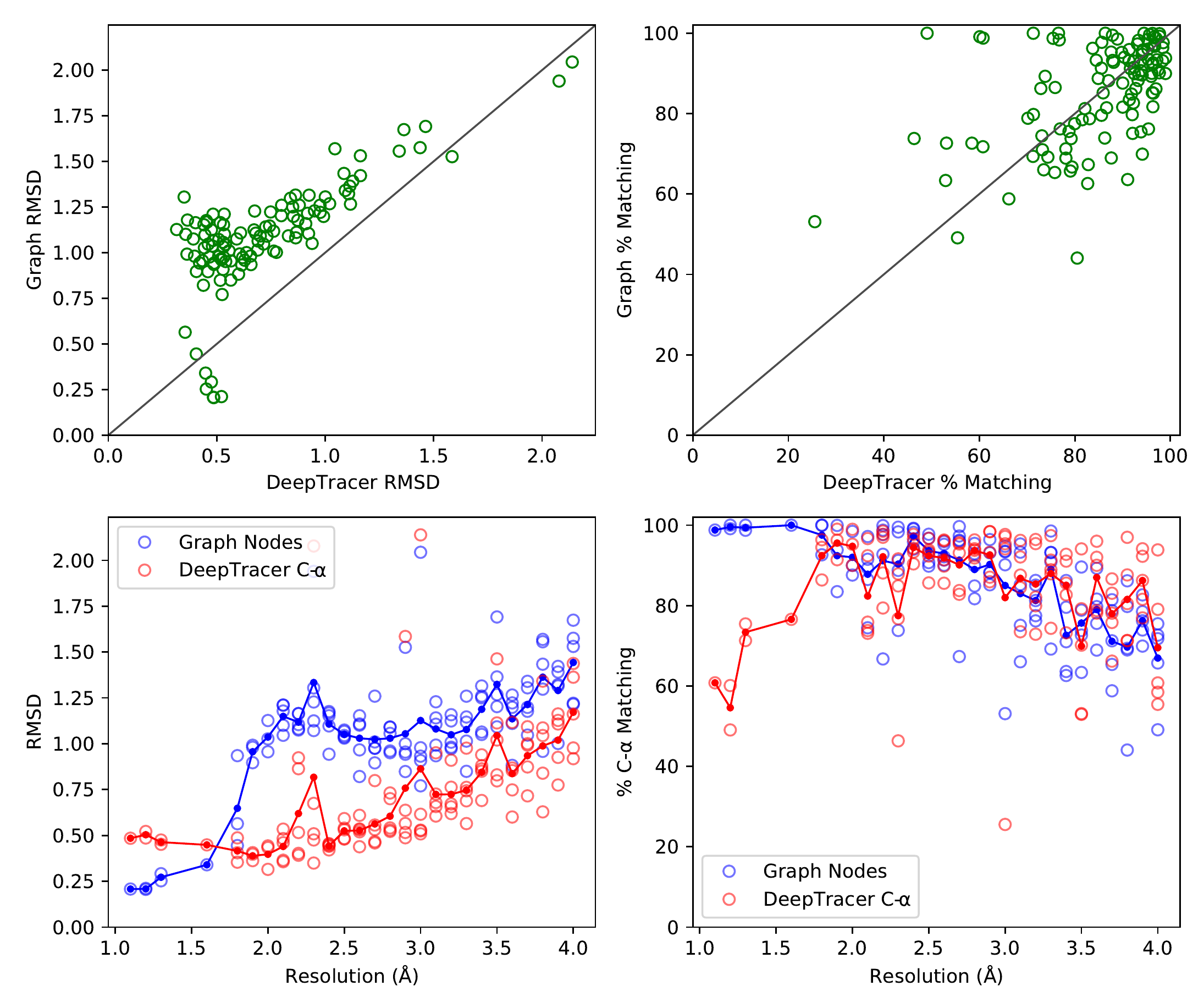}
    \caption{Performance of graph construction compared to predictive output of DeepTracer. The map-wise comparison of the root mean square deviation (RMSD) of nodes and C-$\alpha$ predictions (upper-left), map-wise comparison of the percent of the deposited structure's C-$\alpha$ atoms that have a node or prediction within 3 \si{\angstrom} (upper-right), and the same values but plotted against the corresponding resolutions (lower-left and lower-right respectively).}
    \label{fig:deeptracer_compare}
\end{figure}

The results of node sensitivity to residue locations are competitive when comparing against the C-$\alpha$ predictions of the tool DeepTracer (Figure \ref{fig:deeptracer_compare}). DeepTracer\cite{pfabDeepTracerFastNovo2021} is a method for \textit{de novo} protein structure prediction from high resolution cryo-EM maps that uses a U-Net\cite{ronnebergerUNetConvolutionalNetworks2015} deep convolutional network as the basis for amino acid residue location and type annotations. While the methods are not exactly similar in output, the comparison provides context to the sensitivity metrics of the density graphs. By assuming every node is a potential C-$\alpha$ atom, the overall matching percentage of output to C-$\alpha$ locations is $85.3\%$ and $85.0\%$ for the graphs and DeepTracer predictions respectively. Though the RMSD values of the density graph nodes are worse than the C-$\alpha$ predictions of DeepTracer overall, an average of $1.09$ \si{\angstrom} against $0.72$ \si{\angstrom}, the graphs in the atomic resolution range outperform the output of DeepTracer.

\section{Discussion}
The increasing deposition of maps at high resolutions is a promising sign for the use of our data format and graph-based interpretation in a system of producing atomic structure and type. The density graphs best characterize atomic locations from cryo-EM maps at atomic resolution, and we are unaware of an automated method that performs as well for finding atoms in maps of this resolution. The main significance of this method over simpler methods, such as a simple high-pass filter and clustering method, is that it handles variations of density values present in experimental cryo-EM maps. Visually inspecting the maps shows that the threshold for a high-pass filter that is high enough to capture individual atoms is too high to capture atoms in side chains and on the periphery of the protein (Figure \ref{fig:threshold_problems}a). Our method is able to find density peaks due to the superior interpolation performance of the underlying neural cryo-EM format.

With the cryo-EM density graph creation method applied to 115 experimental cryo-EM maps, it has the additional challenge of dealing with the noise and artifacts present in experimental maps. In order to support dense point detection for both atomic and near-atomic resolution cryo-EM maps, our method does not discriminate between dense points that are relatively close together. Additionally, with our generic initial threshold value calculation, the seed points of some maps may either correspond to noisy regions or may not cover the entire imaged structure. While the specificity of density graphs created from atomic resolution cryo-EM maps is high, over the full range of high resolution maps, the specificity nodes correlated to amino acid residues is relatively low. The rightmost plot in Figure \ref{fig:density_graph_results} shows this with an average of $61.9\%$ of density graph nodes within 3 \si{\angstrom} of any C-$\alpha$ atom, and the variance of this specificity metric is high for density graphs based on near-atomic resolution cryo-EM maps.

\begin{figure}
    \includegraphics[width=0.6\linewidth]{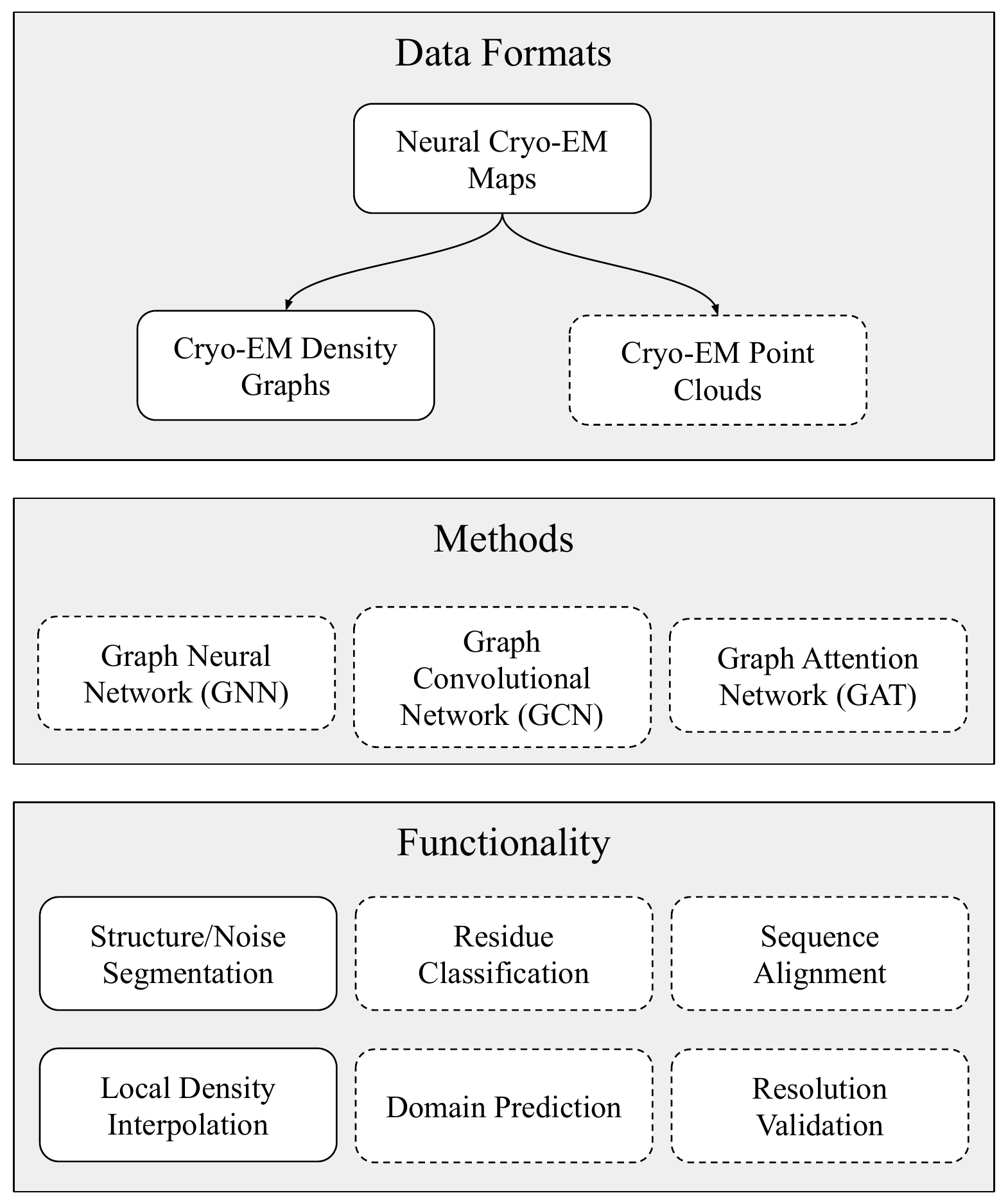}
    \caption{Potential downstream applications of the neural cryo-EM format with items shown in this paper having solid edges. The format itself may be extended to create density graphs, as we have in this paper, or other structures such as a point cloud. Neural cryo-EM maps may also be used directly for tasks such as local density interpolation. Using graph-based machine and deep learning, the graphs created from the neural cryo-EM format may be used to predict individual amino acid type, using maps of atomic resolutions, or other predictive functions depending on the resolution of the original map.}
    \label{fig:downstream_uses}
\end{figure}

As previously stated, experimental noise is a contributor to poorer specificity, and it may be mitigated by a more dynamic or user-influenced initial map threshold. We replaced the generic seed-point threshold with the author-recommended contour value, normalized to the scale of the neural cryo-EM data, as the seed point threshold for a density graph created from EMD-10815. This graph had a high sensitivity with $99\%$ of C-$\alpha$ atoms matching with a node, but was the worst scorer in specificity, showing only $3\%$ of nodes within 3 \si{\angstrom} of a C-$\alpha$ location. This suggested the presence of extensive experimental noise at the initial contour, and, indeed, the re-creation of the graph with the author-recommended threshold resulted in an increase of specificity measurement to $56\%$ while maintaining a $99\%$ sensitivity to C-$\alpha$ locations. Density graphs with low sensitivity were, however, not improved by manually adjusting the seed-point threshold to author-recommended values.

It is also important to highlight that the basis for the cryo-EM density graphs is density peaks, whose meaning changes depending on the resolution of the original cryo-EM map, and this fluidity of context affects the calculation of sensitivity and specificity to concrete locations. Cryo-EM density graphs are inherently descriptive of molecular structure but lack the ability to discern atomic types. We argue that our graph-based interpretation of cryo-EM maps, based on the neural cryo-EM format, is adequate for inclusion in workflows that allow for tailored user interaction in order to provide a boost in node specificity toward molecular structures and type annotation. Fully automated molecular description pipelines may still benefit from the our graph format. Given the demonstrated sensitivity, the cryo-EM density graph is suitable as a pre-processing or initial step in an ensemble of other predictive and refinement methods that operate on graph data structures that may further derive macromolecular context. Figure \ref{fig:downstream_uses} illustrates the potential downstream uses of the neural cryo-EM map and density graph.

\section{Conclusion}
The neural cryo-EM format offers superior interpolation performance compared to conventional linear interpolation, effectively capturing non-linearity of the optimal cryo-EM map imagery across the range of high resolutions. The ability to interpolate cryo-EM map data is preserved in experimentally produced maps as well, validated by using the interpolation to detect locations of density peaks, which correlate to atomic locations. Despite no additional predictive or refinement methods, graphs created from the neural cryo-EM map and detected dense points show similar residue sensitivity to DeepTracer and greatly surpass it for maps of atomic resolution. Density graphs created for cryo-EM maps of atomic resolution cover over 85\% of all atoms in the structure, with the specificity of the nodes equally as high.

The graph data format is especially intriguing with regards to protein representation and structure determination. While our graph node and edge feature vectors are simply spatial data and density data, the high sensitivity of the node placement shows that they may serve as the basis for further predictive methods that use graphs, such as graph convolutional networks. The creation of graphs based on cyro-EM data may facilitate the combination with and integration of methods generally used in other areas of protein structure prediction, such as domain prediction and sequence alignment. As shown by the relatively low specificity and highly contextual nature of our cryo-EM density graph nodes, opportunities to improve our graph format exist, largely in the area of handling noise present in the experimental maps.

As the resolution of cryo-EM maps continues to be driven higher, the demonstrated ability of this format to capture and detect dense locations becomes important to driving future automated methods for determining structure. We chose to implement a graph interpretation of cryo-EM maps, however, due to the nature of the underlying neural cryo-EM format, graphs are not the only possible extension. The accurate interpolation and ability to sample a continuous spatial region may also be used to create other interpretations, such as point clouds. Additionally, as the cryo-EM data is not constrained by voxels and instead represented by neural networks, the format opens the possibility of being integrated into advanced machine and deep learning systems.

\section*{Acknowledgements}
This material is based upon work supported by the National Science Foundation under Grant No. 2030381 and the graduate research award of Computing and Software Systems division at University of Washington Bothell to D.S.. Any opinions, findings, and conclusions or recommendations expressed in this material are those of the authors and do not necessarily reflect the views of the National Science Foundation.

\section*{Abbreviations}
\begin{itemize}
    \item Cryo-EM: Cryogenic Electron Microscopy
    \item 3DEM: Three Dimensional Electron Microscopy,
    \item NMR: Nuclear Magnetic Resonance,
    \item CASP: Critical Assessment of protein Structure Prediction,
    \item GCN: Graph Convolutional Network,
    \item PPI: Protein-Protein Interaction,
    \item SIREN: SInusoidal REpresentation Network,
    \item MAE: Mean Absolute Error,
    \item RMSD: Root Mean Squared Deviation,
    \item C-$\alpha$: Carbon Alpha atom,
    \item PDB: Protein Data Bank,
    \item GPU: Graphics Processing Unit
\end{itemize}

\section*{Competing interests}
The authors declare that they have no competing interests.

\section*{Authors' contributions}
D.S. designed research; N.R. performed research;  N.R. and D.S. analyzed data; N.R. wrote the paper; and D.S. revised the paper.

\bibliographystyle{unsrt} 
\bibliography{references}

\end{document}